\documentclass[ preprintnumbers,A4paper,twocolumn,pra]{revtex4}
\usepackage{graphicx}

\newcommand{\ket}[1]{| #1 \rangle}

\begin{document}

{\Large
\centerline {\bf Quantum Computing with an}
\centerline {\bf `Always On' Heisenberg Interaction}
}
\bigskip
\centerline {{\bf Simon C. Benjamin$^{1,2}$ and Sougato Bose$^{3,4}$}}
\smallskip
{\footnotesize
\centerline {$^1$Ctr. for Quantum Computation, Univ. of Oxford, {\scriptsize OX1 3PU}, UK.}
\centerline {$^2$Dept. of Materials, Parks Road, Univ. of Oxford,  {\scriptsize OX1 3PH},  UK.}
\centerline {$^3$Inst. for Quantum Information, California Inst. of Technology,}
\centerline{ Pasadena,  {\scriptsize CA 91125}, USA.}
\centerline {$^4$Dept. of Physics and Astronomy, University College London,}
\centerline{Gower St., London {\scriptsize WC1E 6BT}, UK.}
}
\bigskip

{\bf Many promising schemes for quantum computing (QC) involve switching 
`on' and `off' a physical coupling between qubits. This may prove 
extremely difficult to achieve experimentally. Here we show that 
systems with a {\em constant} Heisenberg coupling can be 
employed for QC if we actively `tune' the transition energies of 
individual qubits. Moreover we can {\em collectively} tune the qubits to obtain 
an exceptionally simple scheme: computations are controlled via a 
single `switch' of only six settings. Our schemes are applicable to a 
wide range of physical implementations, from excitons and spins in 
quantum dots through to bulk magnets.}

\bigskip
Quantum computing could in principle be performed by a one-dimensional
array of simple systems, such as single electron spins, coupled via
the Heisenberg (`exchange') interaction
\cite{DiVincenzo1,kane,spinResTrans}. Elegant schemes exist whereby this interaction alone generates all
the {\em gates}, or elementary operations on qubits, required for computation
\cite{3qubitExchangeOnly,myABqubitPaper}. It is also known that it can suffice to control the qubits
collectively \cite{ababPRL}. However, all
these schemes require the experimentalist to control the
magnitude of the Heisenberg interaction - effectively to be able
to switch it `on' and `off'\cite{jos}. A typical idea for achieving this is to somehow dynamically manipulate the
wavefunction overlap between neighboring qubits. This appears feasible, but {\em highly} challenging.
Recently Zhou {\em et al.}\cite{zhou} have explored a possible means of avoiding this switching. They observe that the Heisenberg interaction can be effectively negated by inserting EPR spin pairs between the qubits in a (necessarily) two-dimensional architecture. The approach is conceptually rather beautiful, but from a practical point of view it is complex in terms of the physical arrangement of qubits, the initialization, and the steps involved in generating gates. Here we take an entirely different approach and demonstrate that an `always-on' interaction can suffice even in a generic one-dimensional array. Our gate procedure is very simple and can support additional features in suitable systems: the entire device can be controlled {\em without} local manipulation of any kind, and the Zeno effect can be harnessed to reduce errors. 

For convenience of
exposition, we will use the terms `spin' and `Zeeman energy' to
refer to our generic two-state systems and their level splitting. Consider a linear chain of $N$ spins, with a Hamiltonian
${\hat H}={\hat H}_{\rm Zeeman}+{\hat H}_{\rm int}$ where: 
$${\hat H}_{\rm Zeeman}=\sum_{i=1}^NE_i(t){\hat \sigma_i^Z}, \ \ \ \ {\hat H}_{\rm int}=J\sum_{i=1}^{N-1}{\hat {\underline \sigma}}_{i}.{\hat {\underline
\sigma}}_{i+1}.$$ 
\noindent  Here $\hbar=1$ and subscript $i$ denotes an operator
acting in the subspace of the $i^{th}$ qubit. \{${\hat \sigma}^X$, ${\hat \sigma}^Y$, ${\hat \sigma}^Z$\} are the Pauli matrices,
and ${\underline {\hat \sigma}}\equiv{\underline i}{\hat
\sigma}^X+{\underline j}{\hat \sigma}^Y+{\underline k}{\hat
\sigma}^Z$. Zeeman energies $E_i$ may vary with time, but the interaction couples all
nearest-neighbors with a common magnitude and is {\em constant}.
We exploit the well-known
observation that when the Zeeman energies vary to the extent that
$|E_i-E_{i+1}|\gg J$, then the interaction tends to an
{\em effective} Ising form \cite{CiS}: $ {\hat H}_{\rm int}\approx J \sum 
{\hat \sigma}^Z_i {\hat \sigma}^Z_{i+1}$.
\begin{figure}[!t]
  \begin{center}
    \leavevmode
\resizebox{7.8 cm}{!}{\includegraphics{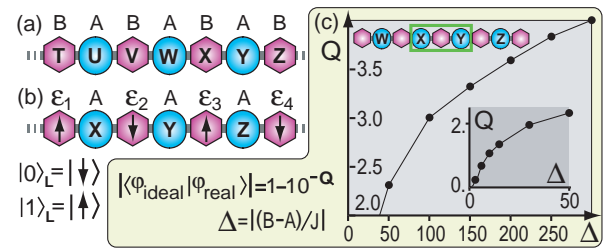}}
\end{center}
\vspace{0.2cm}
\caption{Strategies for implementing QC on a 1D chain. Blocks represent individual spins and qubits are denoted by letters $T,U$..$Z$. Fixed Zeeman energies are denoted by $A$ \& $B$, tunable Zeeman energies by $\epsilon_i$. Here we assume an independent mechanism exists for single qubit gates. If these are much faster than $1/J$, then one simply adopts the trivial scheme shown in (a). Qubits are placed on adjacent cells and will suffer continuous phase gates with their neighbors, but techniques developed for NMR QC\cite{NMRpaper} can be employed to actively negate this via fast rotations. However universal single
qubit gates in solid state systems are usually slow\cite{3qubitExchangeOnly}
compared to $1/J$. In this case we would choose to place qubits only on alternate spins (b), with
intervening spins in definite classical states. The Ising interaction
is then entirely negated, and two-qubit operations are achieved by `tuning' a spin's Zeeman energy, as described in the text. Part (c)  displays data from a numerical analysis of the process: a gate on $X$ and $Y$ is achieved by tuning $\epsilon_2$ in the nine spin section shown. Plots show worst case defects over the 16 possible basis states. Phase noise (not shown) was always smaller. }
\label{figure1}
\end{figure}

The choice of scheme for performing QC on such systems depends on the available
experimental abilities:

{\bf (1) Ability for universal single qubit gates.}
Suppose that a mechanism exists whereby general rotations of
individual spins can be performed (essentially the same physical starting point assumed by Zhou {\em et al}\cite{zhou}). If such rotations are extremely fast, then there is an immediate solution (Fig. 1a). Alternatively, Fig. 1b shows the approach when fast tuning of Zeeman energies is possible\cite{fastTune}, but the additional manipulation(s) used to compose universal single qubit gates are not rapid.
We separate the qubits to negate their continuous residual Ising interaction; a
two-qubit gate must then involve temporarily altering the pattern of Zeeman energies. Consider
a section of the array with the initial Zeeman pattern $BABAB$ and
containing two qubits represented by the states of the $A$ spins
(as the five leftmost spins in Fig 1b). Referring to these
spins by the numbers $1$..$5$, assume that the
outer spins $1$ and $5$ are in state $\ket{\uparrow}$, and that the
central spin $3$ is $\ket{\downarrow}$. We will show that a gate can be achieved by
tuning {\em only} the Zeeman energy of spin $3$, which we
will denote $\epsilon(t)$. Since the Zeeman energies of spins $1$
and $2$ will remain far out of resonance, the interaction between them will
remain of the form $J{\hat \sigma}_1^Z{\hat \sigma}_2^Z$. Then spin
$1$ remains in state $\ket{\uparrow}$ throughout,
effectively producing a shift of $+J$ to the Zeeman energy of spin $2$. The
same holds for spins $4$ and $5$, so we can describe the non-trivial dynamics
of the five spins via a three spin Hamiltonian:
$$
{\hat H}=(A+J)({\hat \sigma}_2^z+{\hat \sigma}_4^z)+\epsilon(t){\hat \sigma}_3+J({\hat {\underline \sigma}}_2.{\hat {\underline \sigma}}_3 + {\hat {\underline \sigma}}_3.{\hat {\underline \sigma}}_4)
$$
Suppose that at $t=0$ we move abruptly from the passive state
$\epsilon=B$ to the perfectly resonant case $\epsilon=(A+J)$.
Then the two qubits $X$ and $Y$ will `spread' over all three cells. However at a time $t_r=\hbar/(6J)$, spin $3$ returns to its
$\ket{\downarrow}$ state and an unitary transformation ${\hat G}$ is achieved
between $X$ \& $Y$. In their computational
basis \{$\ket{00}_{24},\ket{01}_{24},\ket{10}_{24},
\ket{11}_{24}$\} we find \cite{labFrame}

${\hat G}=\left(
\begin{array}{cccc}
1 & 0 & 0 & 0 \\
0 & W & i{\sqrt 3}W & 0 \\
0 &  i{\sqrt 3}W & W & 0\\
0 & 0 & 0 & 1
\end{array}
\right) $
where $W={1\over 2}e^{i \pi/3}$.

\noindent This is an entangling gate and it is simple to use
established formalisms\cite{gatePaper,nielsen} to
generate a CNOT using 4 applications of ${\hat G}$. Therefore tuning
the Zeeman energy of `barrier' spins is adequate, in
combination with single-qubit gates, to efficiently implement
quantum algorithms. In a large array of spins, we can apply this
process independently to barrier spins at various points -
therefore we have complete parallelism in this architecture. This
is a requirement for full quantum error correction \cite{steane}. In this scheme and the following ones, barrier spin initialization can be achieved by relaxing to the spin-polarised ground state followed by selective spin rotations, either via local gates or frequency selective global pulses.

{\bf (2) No single qubit ability.} Suppose that we cannot perform general rotations on individual spins (we can {\em only}
tune their Zeeman energies). Then we adopt the architecture shown in Fig 2a. 
The passive state of the device now has a sequence of Zeeman energies
$ABCABC..$ with $C-B\gg J$ and  $B-A\gg J$. (Other
patterns such as $ABABAB..$ may still suffice \cite{longPaper},
but $ABC$ is convenient for the purpose of exposition.) Qubit representation is as specified in Fig 2a, and the mechanism for single-qubit gates is illustrated in Fig 2b. Fortuitously this encoding constitutes a subspace that protects against long wavelength phase noise. This is the prevalent noise in many systems, as with low temperature phonons in the solid state for example. The one- and two- qubit gates defined in Fig. 2 respect the constraint that there is exactly one $\ket{\downarrow}$ spin among the three associated with each qubit. 

\begin{figure}[!t]
  \begin{center}
    \leavevmode
\resizebox{6.3 cm}{!}{\includegraphics{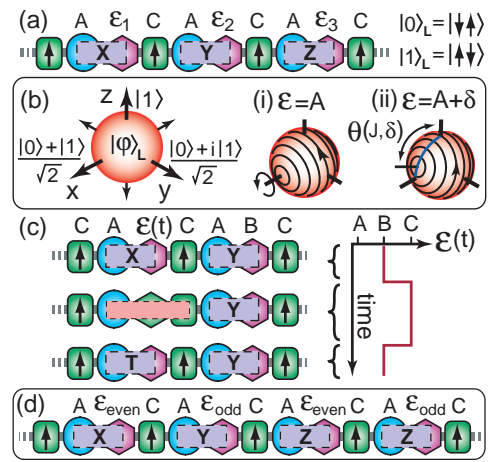}}
\end{center}
\vspace{0.2cm}
\caption{Strategies for implementing QC when no independent mechanism for single qubit gates exists; these are now synthesised via Zeeman tuning alone. This requires an encoding (a) of two spins per qubit, with a third acting as a barrier. All gates can be implemented purely by tuning the
Zeeman energy of the qubit that initially has energy
$\epsilon=B$. For single qubit gates: If we abruptly tune this energy to
$\epsilon=A$, perfectly matching the energy of its neighbor,
then the logical qubit represented by this pair will experience \cite{labFrame} a
continuous rotation given by $\exp(i k t {\hat \sigma}_x)$ as shown in (b)(i). Tuning to an energy $\epsilon=A+\delta$ will
produce a rotation about an axis in the z-x plane, the axis being
determined by the ratio $J$ to $\delta$, as depicted in (b)(ii). Such rotations
can synthesize any one-qubit gate. For a two-qubit gate we employ the process shown in (c): we set the energy $\epsilon$ of our tunable spin to a value near $C$. This effectively allows qubit $X$ to `spread' onto the barrier spin, where it experiences a conditional phase gate due to the proximity $Y$. Part (d) shows an architecture for the case where Zeeman energies cannot be tuned independently for nearby spins. QC can still be achieved, by collectively tuning one of the two subsets with energies $\epsilon_{even}$ and $\epsilon_{odd}$. }
\label{figure2}
\end{figure}

\noindent To perform a two-qubit gate, we allow a qubit to `spread' onto the `barrier' spin as shown in Fig 2c.
Our complete process must of course return the `barrier' back to the definite state
$\ket{\uparrow}$. By modelling a four-spin section $A\epsilon CA$, we find that this can be achieved
by choosing $\epsilon=C+J$: a suitable `revival' of the barrier spin then occurs at time
$t_r=(\pi/\sqrt 5)(\hbar/J)$. It is easy to show that the resulting unitary
transformation ${\hat K}$, along with two suitable single-qubit gates \cite{longPaper},
generates the transformation
$$
{\hat M}=\left( {\hat Q1} \otimes {\hat Q2}\right).{\hat K}=
\left(
\begin{array}{cccc}
1 & 0 & 0 & 0 \\
0 & 1 & 0 & 0 \\
0 & 0 & 1 & 0\\
0 & 0 & 0 & e^{{-i\pi}\over \sqrt 5}
\end{array}
\right)\ \ $$

\noindent in the basis of qubits $X$ and $Y$. We can then use established formalisms\cite{gatePaper} to generate a CNOT gate using two applications of ${\hat M}$. We later note that this gate has some advantages over the one employed in our $1^{st}$ architecture.

{\bf (3) No ability for local gates.}
Until now we have assumed that the experimentalist can tune the
spins {\em independently} from one-another. Even
this requirement can be dispensed with, using a variant of the method defined in Ref.\cite{ababPRL} (a descendant of Lloyd's  global control scheme \cite{sethSci}).
  Consider the architecture of the previous section in which one in every three spins is `tunable'. Now notionally
divide those spins into two groups, the `odd' and `even' groups, in an alternating pattern (Fig 2d). Introduce the dramatic simplification that all spins within the odd group have the same energy $\epsilon_{odd}$, and similarly for the even group. Now suppose that we permit ourselves to tune
$\epsilon_{even}$, $\epsilon_{odd}$ through a sequence of values
{\em always} respecting the constraint $|\epsilon_{odd}-C| \gg J$.
By the results of the previous section, we know that this will allow us to perform any single qubit gates on the corresponding {\em
odd} and {\em even} qubits, and to produce our phase gate ${\hat M}$ between
each {\em even} qubit and the {\em odd} qubit to its right. This process, together with the complementary  process (under constraint $|\epsilon_{even}-C| \gg J$), meets the fundamental conditions in Ref.\cite{ababPRL}. In this way we can immediately translate the protocol defined there to the present scheme.
Universal QC (including error correction\cite{steane}) on our entire multi-qubit device is thus governed via {\em
global} experimental parameters: $\epsilon_{odd}$ and
$\epsilon_{even}$. Moreover these parameters need only assume
certain fixed values, given that the duration is a continuous
variable: for one-qubit gates, $A$ \& $A+J$ (say), and for a two-qubit gate,
 $C+J$.
 It follows that only six specific pairs of values for $\epsilon_{even}$,$\epsilon_{odd}$ suffice (Fig.3 left). 
 
\begin{figure}[!t]
  \begin{center}
    \leavevmode
\resizebox{7.8 cm}{!}{\includegraphics{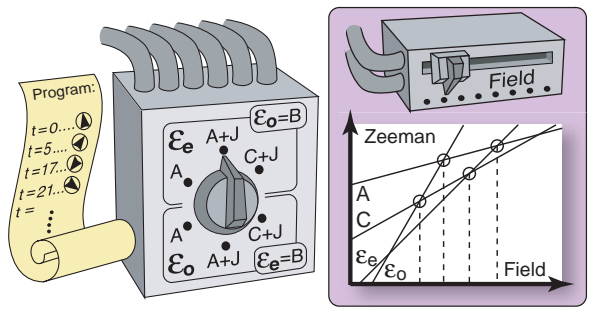}}
\end{center}
\vspace{0.2cm}
\caption{Left: Cartoon emphasizing that, in our $3^{rd}$ architecture, the entire computation on all qubits can be implemented simply by switching between 6 settings. In practice such switching would be of course be performed by a conventional computer (not manually!), and moreover one would require additional setting(s) for measurement. Right: in certain physical systems, the switch can be equivalent to adjusting just a {\em single} global parameter.}
\label{figure3}
\end{figure}

 We have used the term `global' even though the process does require differentiation on the local scale between the regular sets of `even' and `odd' spins (e.g. via a single electrode running the length of the device, patterned at the local scale \cite{ababPRL}). But even this requirement can be dispensed with in suitable systems, to yield {\em pure} global control. The necessary physical property is illustrated on the right side of Fig. 3: the Zeeman energies of certain classes of spin must intersect as some external parameter (typically a field strength) is swept. Since multiple Zeeman energies then change simultaneously, unwanted phase shifts would occur during two-qubit operations - but these can be compensated for in subsequent steps. A possible physical realization is noted below.  

To meet the ultimate goal of full scale QC, one must suppress all operational error rates sufficiently for them to be handled by general error correction protocols\cite{steane}.  We now review the potential error sources that are intrinsic to our schemes. {\bf Imperfectly localized tuning of Zeeman energies}: In reality nearby spins may be effected to some degree. Fortunately all our schemes are very robust against this effect. The two-qubit gates rely on being able to bring two spins into resonance, which is possible even if the second spin is experiencing a small tuning effect. At worst one would simply generate an easily-corrected phase shift. {\bf Imperfect gate operations: } All QC proposals inevitably demand exquisite control of their physical gate processes (to within error correction thresholds). For our schemes this means precise timing of the spin resonance periods. An advantage of our approach is that there is a simple tactic to make this goal more achievable: we can `put to work' our redundant barrier spins via the Quantum Zeno Effect\cite{zenoEffect}. If we repeatedly collapse the state of the barrier spins to their $\ket{\uparrow}$, $\ket{\downarrow}$ basis, on a time scale short compared to the rate at which they would accumulate errors, then we can actually {\em suppress} that accumulation. Note that we use the term `collapse' rather than `measurement' to emphasize that the phenomenon does not require one to detect the outcome. For maximum efficiency the process should be performed simultaneosuly for all barrier spins. The process fails if a spin ever collapses to the `wrong' state - but the {\em total} probability of such an event vanishes with increasing frequency of collapse. Therefore the ideal would be to collapse the barrier spin wavefunctions after {\em each} gate operation (although never during an operation of course). Later we outline some systems that could support this idea. We emphasize that this exploitation of the Zeno effect is not a requirement of our schemes - we are merely observing that {\em if} the phenomenon is supported by a physical system then we can make good use of it! In cases where it is not possible then it may be desirable to opt for the type of two-qubit gate employed in schemes $2$ \& $3$: an imperfect operation there would not generate three-qubit correlations. {\bf Irregularity in physical separations, interaction strengths or susceptibility to Zeeman tuning}: these could only be tackled by `calibrating' the system and tailoring the set of Zeeman shifts to each spin uniquely. Again the form of two-qubit gate in architecture $2$ is robust, since this does not rely on a symmetry of two interactions. However the non-local addressing in scheme $3$ cannot accommodate inhomogeneities in this way, therefore it is only suited to very regular structures (e.g. periodic molecular systems, or atomically accurate quantum dot arrays). {\bf Finite value of $\Delta$}: the importance of this energy ratio is shown in Fig.1c. For large values of $\Delta$ the gate is near perfect, but the fidelity falls with $\Delta$ and below $10$ it rapidly becomes unusable, except perhaps for initial `proof in principle' experiments. Therefore ideal physical implementations will be those in which strong tuning of the Zeeman energy is possible.

We will now highlight a few realizations. Our schemes are relevant to {\em all} proposals for quantum computing that involve a
Heisenberg interaction and can support tuning of transition energies. It is natural to first consider `true' spin systems, e.g. single electron arrays. These are often discussed as potential quantum computers; typical proposals involve a mechanism for switching the interaction and a second independent mechanism for performing single qubit gates. Our schemes allow one to dispense with the former and retain only the latter. To exploit the Zeno effect, one could employ the Pauli blockade phenomenon: a suitable optical pulse can  {\em conditionally} create an exciton (a bound electron-hole pair) in the region of a preexisting electron (the qubit) {\em depending} on its state. A previous QC proposal makes sophisticated use of this idea \cite{Pazy} but here we exploit it very crudely: merely by allowing the exciton to dissipatively decay (or to relax), one would indirectly collapse the state of the electron spin.

Our schemes are also relevant to a different class of system that operates (and decoheres) on a far more rapid time scale: pure exciton computing. Exciton lifetimes are very short, however the gates can be ultrafast (picoseconds) so that a large number of operations could be performed within that lifetime.
In typical exciton QC schemes the up/down pseudo-spin states are the presence/absence of an exciton on a quantum dot (QD), thus our `Zeeman' energy would correspond to the exciton creation energy. This could be tuned either by shifting the exciton localization between regions of different band gap (somewhat analogously to Ref.\cite{spinResTrans}) or via the quantum confined Stark effect. The latter is expected\cite{ShengPRL}  to be very strong in double-dot structures: of order 100 meV for achievable fields. The coupling strength $J$ between two separated pairs may be a fraction of 1 meV, leading to a suitably large ratio $\Delta$.  Both the D.C. and A.C. Stark effects are relevant - the latter could permit `all-optical' control. Moreover the Stark effect is seen in many other quantum systems (including molecular structures) and could allow them to be similarly exploited. Furthermore, since the creation energy is non-zero at zero field, the Stark effect could in principle support the `pure' global switching illustrated in the right side of Fig.  3. Exciton systems can also provide sufficiently rapid wavefunction collapse for our Zeno exploitation, e.g. via a laser tuned to generate an {\em excited} exciton state with rapid (picosecond) intra-band relaxation. The required frequency may be distinct for barrier spins versus qubit-bearing spins (given that the barrier spins must have a distinct exciton creation energy) - if so then one could simultaneously collapse all barrier spin states with a global pulse.

Thus it appears that several of the phenomena associated with excitonic systems may be well suited to our purposes. Looking beyond such systems, we speculate that the minimal demands of our $3^{rd}$ architecture may introduce the possibility of QC to new classes of system. For example, in 1D Heisenberg magnets such as KCuF$_3$\cite{schultz},  the effect of coupling between
two chains can be replaced by an effective
inhomogeneous magnetic field on one of the chains \cite{schultz}.
Zeeman tuning might then be accomplished by controlling
the distance and alignment of one 1D spin chain with respect to another. 

SCB wishes to acknowledge support from a Royal Society URF, and from the Foresight LINK project  ``Nanoelectronics at the Quantum Edge''. SB acknowledges the NSF under Grant Number EIA-00860368.


\begin{references} 

\bibitem{DiVincenzo1} D. Loss \& D. P. DiVincenzo, Phys. Rev. A {\bf 57}, 120 (1998).
\bibitem{kane} B. E. Kane, Nature {\bf 393}, 133 (1998).
\bibitem{spinResTrans} R. Vrijen {\em et al}, Phys. Rev. A {\bf 62}, 012306 (2000).
\bibitem{3qubitExchangeOnly} D. P. DiVincenzo {\em et al}, Nature {\bf 408}, 339 (2000).
\bibitem{myABqubitPaper} S. C. Benjamin, Phys. Rev. A {\bf 64}, 054303 (2001).
\bibitem{ababPRL} S. C. Benjamin, Phys. Rev. Lett. {\bf 88}, 017904 (2002).
\bibitem{jos} This requirement exists also in superconducting qubit schemes; see e.g. Y. Makhlin {\em et al},  Nature {\bf 398}, 305 (1999).
\bibitem{zhou}X. Zhou {\em et. al}, Phys. Rev. Lett. {\bf 89}, 197903 (2002).
\bibitem{NMRpaper} J. A. Jones, \& E. Knill, J. Mag. Res. {\bf 141}, 322 (1999).
\bibitem{CiS} P.W. Anderson, {\em Concepts in Solids}, Addison Wesley, Redwood City 1963.
\bibitem{fastTune} It is likely that this condition of abrupt tuning can be relaxed - one could make $\epsilon(t)$ a smooth function and numerically search for suitable barrier spins `revival' points.
\bibitem{labFrame}Throughout we use the lab frame (i.e. pure Schrodinger picture), so that other `passive' qubits perform $\sigma_Z$ rotations at a common fixed rate.
\bibitem{gatePaper} M. J. Bremner  {\em et al.}, http://arxiv.org/abs/quant-ph/0207072.
\bibitem{nielsen}J. L. Dodd {\em et al}, Phys. Rev. A {\bf 65}, 040301 (2002).
\bibitem{steane} A. M. Steane, http://arxiv.org/abs/quant-ph/0207119. Error correction in our $3^{rd}$ architecture requires the ability to reset specific spins - but this can still be controlled globally (A. Bririd \& S. C. Benjamin [unpulished]).
\bibitem{longPaper} S. C. Benjamin \& S. Bose. (unpublished).
\bibitem{sethSci} S. Lloyd, Science {\bf 261}, 1569 (1993).
\bibitem{zenoEffect} P. Facchi {\em at al}, Phys. Rev. Lett. {\bf 86} 2699 (2001).
\bibitem{Pazy} E. Pazy {\em et al.} preprint  lanl.arxiv.org/cond-mat/0109337.
\bibitem{ShengPRL} W. Sheng \& J. P. Leburton, Phys. Rev. Lett. {\bf 88}, 167401 (2002).
\bibitem{schultz} H. J. Schulz, Phys. Rev. Lett. {\bf 77}, 2790 (1996).

\end{references}
\end{document}